\documentclass[referee,useAMS,usegraphicx,usenatbib]{mn2e}
\usepackage{times}

\newcommand{\Msun}{\ensuremath{\,{\rm M}_\odot}}                        
\newcommand{\Rsun}{\ensuremath{\,{\rm R}_\odot}}                        
\newcommand{\kms}{\,km\,s$^{-1}$}                                       

\title[TYC\,1031\,1262\,1: An Anomalous Cepheid in a double-lined eclipsing binary]
      {TYC\,1031\,1262\,1\thanks{Based on the data obtained at T\"{U}B\.{I}TAK National Observatory}: An Anomalous Cepheid in a double-lined eclipsing binary
      }

\author[ E. Sipahi et al.]
  {E.~Sipahi$^1$\thanks{e-mail:esin.sipahi@mail.ege.edu.tr}, C.~\.{I}bano\v{g}lu$^1$,  \"{O}. \c{C}ak{\i}rl{\i}$^1$, S. Evren$^1$  \\ \\
  $^1$Ege University, Science Faculty, Astronomy and Space Sciences Dept., 35100 Bornova, \.{I}zmir, Turkey\\  }

\begin{document} \maketitle 

\begin{abstract}
Multi-color light curves and radial velocities for TYC\,1031\,1262\,1 have been obtained  and analyzed. TYC\,1031\,1262\,1 includes a Cepheid with a period of 4.15270$\pm$0.00061 days. The orbital period of the 
system is about 51.2857$\pm$0.0174 days. The pulsation period indicates a secular period increase with an amount of  2.46$\pm$0.54 min/yr. The observed B, V, and R magnitudes were cleaned for the intrinsic 
variations of the primary star. The remaining light curves, consisting of eclipses and proximity effects, are obtained and analyzed for orbital parameters. The system consists of two evolved stars, F8II+G6II, with 
masses of M$_1$=1.640$\pm$0.151 {\Msun} and M$_2$=0.934$\pm$0.109 {\Msun} and radii of R$_1$=26.9$\pm$0.9 {\Rsun} and R$_2$=15.0$\pm$0.7 {\Rsun}. The pulsating star is almost filling its corresponding 
Roche lobe which indicates the possibility of mass loss or transfer having taken place.  We find an average distance of d=5070$\pm$250\,pc using the BVR and JHK magnitudes and also the V-band extinction. Kinematic 
properties and the distance to the galactic plane with an amount of 970 pc indicate that it belongs to the thick-disk population. Most of the observed and calculated parameters of the TYC\,1031\,1262\,1 lead to a classification of an 
Anomalous Cepheid.     
\end{abstract}



\section{Introduction}                                                                                                               \label{sec:intro}
Pulsating variables are located in a restricted region, so-called  Cepheid instability strip (IS) in the Hetrzsprung-Russell (HR) diagram. This region includes 
$\delta$\,Scuti, RR\,Lyrae, Population\,I and II\,Cepheids which represent an important testing basis for theories on 
stellar structure and evolution as well as pulsation mechanisms. These variables have very different masses, effective temperatures and 
chemical abundances. The stars in the IS are found not only in a wide range of masses but also found among very young Population\,I type stars to very old  Population\,II type stars. These stars in the IS reveal many properties of the stellar interiors.   

At the beginning of the 20$^{th}$ century Miss Leavitt made a great discovery that there is a relation between the period and absolute magnitude 
of the Cepheid variables in the Small Magellanic Cloud. Three decades later \citet{Baa53} called attention to the difference in the 
period-luminosity (hereafter P-L) relations of the Cepheids in the globular clusters and the classical Cepheids. Thus, the Cepheids are 
divided into two groups which are Population\,I (usually called Type\,I Cepheids) and Population\,II Cepheids (or Type\,II Cepheids), the formers are about 1.5 mag brighter than the latter for the same pulsation periods. Taking into account this difference \citet{Bla54} made a revision on the P-L relation for the classical Cepheids. These variables are still used as distance indicators not only for our galaxy but also for nearby spiral and irregular 
galaxies. While the Type\,I Cepheids are young-disk stars associated with spiral arm and thin-disk populations evolved from main-sequence 
stars with masses from 3 to 15 solar masses, Type\,II Cepheids can be old-disk, thick-disk, or halo stars with masses somewhat smaller than 
the Sun (see \citet {Wal02}, and references therein). Type\,II Cepheids are divided into three subclasses regarding their pulsation periods: BL\,Her stars with a period of $<$ 7\,d, W\,Vir stars with 7$<$  P $<$ 20 d and RV Tau type variables with P$>$  20 d. Recently  \citet{sos08} suggested divisions at 4 and 20 days. Masses and radii for these stars were estimated from their pulsation 
properties. BL Her stars are common in globular clusters and in the disk of our galaxy. \citet{Har81} showed that four of the 12 field 
BL\,Her stars have [Fe/H] values less than -1.5 which were identified as halo objects with a mean distance from the galactic plane of 
1.8\,kpc. In contrary, the 8 stars have metallicities near or above solar values. Their mean distance from the Galactic plane was about 
0.25\,kpc, indicating characteristic of thick or thin-disk population. There are some metal-poor variables with periods between 0.5 and 3 or more days lying above the RR\,Lyrae-type stars which may have entirely different origin as compared to the normal Type\,II Cepheids in the globular clusters. These variables are called as Anomalous Cepheids (hereafter ACs). They are found in the general field, globular clusters and nearby dwarf spheroidal galaxies. Anomalous Cepheids are brighter than the Type\,II Cepheids at fixed color, and hence they follow a different P-L relation \citep{sos08}. \citet{fio06} (and references therein) suggested that the ACs are extension of the Type\,I Cepheids to lower metallicities and masses. Moreover, they predict a mass range for the ACs as 1.9$<$M/M$_{\odot}$ $<$3, which may be a product of a mass about  M$<$ 4 M$_{\odot}$, experienced the helium flash in the past.             
 
Only six Type\,II Cepheids were known to be spectroscopic binaries. TX\,Del, IX\,Cas, AU\,Peg and ST\,Pup are single-lined binaries for which 
spectroscopic observations and the resultant orbits were published. Due to low orbital inclination no eclipses were detected in these 
binaries. Recently, two Cepheid variables in eclipsing binaries, namely TYC\,1031\,1262\,1 and NSV\,10993, were discovered by \citet{Ant07} and \citet{Khr08}.
The pulsation periods are nearly identical with an amount of 4.2\,d, but the orbital periods are about 51 and 40 days. In the second part of the OGLE-III catalog of Variable stars \citet{sos08} presented 197 Type\,II Cepheids of which seven are eclipsing variables and 83 ACs in the Large Magellanic Cloud.   
Since masses and radii for the stars were measured directly from the eclipse light curves and radial velocities, a detailed study of these Cepheids 
in the eclipsing binaries would be well worthwhile. 

The light variability of TYC\,1031\,1262\,1 (ASAS\,J182611+1212.6, 2MASS\,J18261150+1212349, GSC\,01031-01262, V=11$^{m}$.64, B-V=0$^{m}$.77) was discovered 
by \citet{Poj05} during the All-Sky Automated Survey (ASAS-3). The variability of this star was also detected independently by \citet{Ant07} 
on Moscow archive plates in Crimea. They plotted the observations obtained by ASAS-3, photographic magnitudes obtained from the plates taken 
with the 40 cm astrograph in the Crimea and the observations obtained by the Northern Sky Variability Survey (NSVS, \citet{Woz04}) against 
the time and noticed a new variable as a Cepheid with some peculiarities. They started observations of TYC\,1031\,1262\,1 with the 50\,cm 
telescope of the Crimean Laboratory. These CCD observations clearly revealed eclipsing nature of the star with a Type\,II Cepheid component 
in our Galaxy. The pulsation and orbital periods are estimated to be 4.1523 and 51.38\,d, respectively. \citet{Sch09} made VR photometry of Cepheid variable star candidates including TYC\,1031\,1262\,1 in 2005 and 2006. They proposed a pulsation period of about 4.1508\,d and classified it as a Type\,I Cepheid.   
               
In this study we present our multi-color photometric and spectroscopic observations of TYC\,1031\,1262\,1. The main aim of this study is to 
derive the masses and radii of the components. Thus, mass and radius of a Type\,II Cepheid will be revealed, for the first time, directly 
from the spectroscopic and photometric observations. We will also discuss the pulsation characteristics and its place in our Galaxy.

\section{Observations}
\subsection{Photometric observations}
The photometric observations in the wide-band Johnson UBVR system were carried out with the 48\,cm Cassegrain reflecting telescope
and 35\,cm MEADE\,LX200\,GPS telescope at the Ege University Observatory. In the observations with the 48\,cm telescope High-Speed Three-Channel 
Photometer and standard UBVR passbands were used. Thermoelectrically cooled ALTA U+42 2048x2048 pixel CCD camera including BVR passbands was 
attached to Schmidt Cassegrain type Meade telescope. The BVR observations in 2008 were obtained on 63 nights between March\,2 and November\,5. The 
observations in 2009 were obtained on 17 nights between August\,3 and October\,8. GSC\,1031\,193 and GSC\,1031\,1445 are taken as the comparison and 
check stars, respectively. Some basic parameters of the comparison stars taken from the SIMBAD database are listed in Table\,1. Although the program 
and comparison stars are very close in the sky, differential atmospheric extinction corrections were applied, especially for the observations obtained with the 48\,cm 
telescope. While all the program stars were observed simultaneously with the 35\,cm telescope, each star was observed successively with the 48\,cm 
telescope, i.e. in different times. The atmospheric extinction coefficients were obtained from observations of the comparison stars on each 
night. Moreover, the comparison stars were observed with the standard stars in their vicinity and reduced differential magnitudes, in the sense
variable minus comparison, were transformed to the standard system. The standard stars are chosen from the lists of Landolt (1983, 1992). Heliocentric 
corrections were also applied to the times of the observations.

In Fig.\,1 we plot the U-,  B-, V- and R-passband observations versus the pulsation period of the Cepheid variable. As it is seen the dominant light variations are 
originated from the pulsation. The below shifted observations correspond to the eclipses. The standard deviations of each data point are about 0.05, 0.03, 0.01, and 0.01 mag in U,  B, V and R passbands, respectively. The observational data can be obtained from the authors. 
    
\begin{figure}
\center
\includegraphics[width=12cm,angle=0]{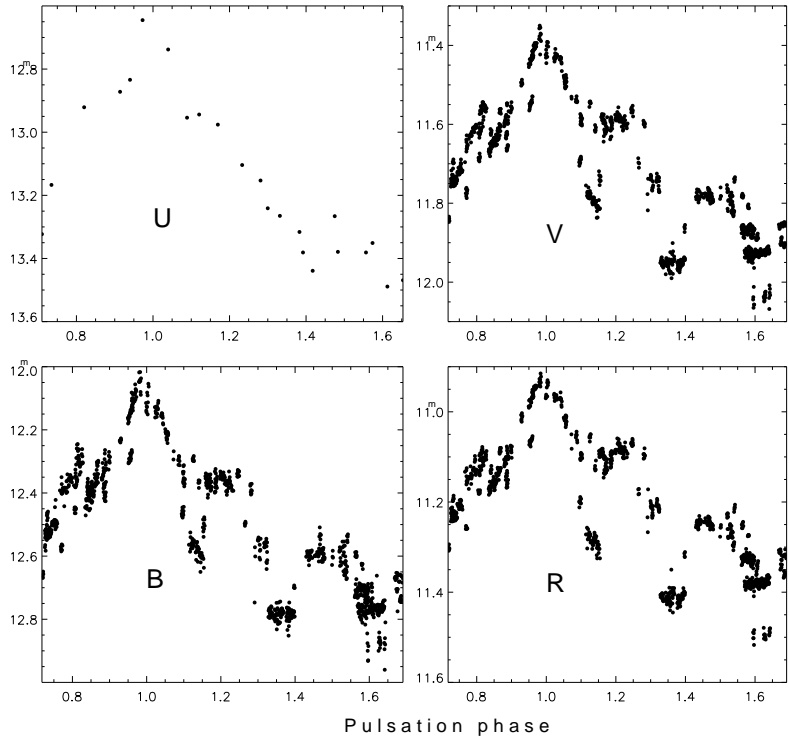}\\
\caption{The U-, B-, V-, and R-passband light curves for TYC\,1031\,1262\,1. The ordinates and apsis are the standard magnitudes and the pulsation phases. } 
\end{figure} 

\begin{table*}
\centering
\footnotesize
\begin{minipage} {12 cm}
\caption{The coordinates, apparent visual magnitudes and the colors of the stars observed.}
\begin{tabular}{|l|c|c| c| c|}
\hline
\hline
Star	& $\alpha$ 	&$\delta$    &V (mag)	&	B-V(mag)	\\
\hline					
TYC\,1031\,1262\,1  & 18$^h$ 26$^m$ 11.5$^s$	& 12$^{\degr}$ 12$^{'}$ 34$^{''}$.8     & 11.64	& 0.77		\\
GSC\,1031\,193 	& 18 26 23.7  & 12 15 47.2     &  10.99	&  0.66  \\
GSC\,1031\,1445	& 18 26 04.9  & 12 07 11.6     &  10.44   &  0.63	\\
\hline
\end{tabular}
\end{minipage}
\end{table*}
\smallskip

\subsection{Spectroscopic observations}
Optical spectroscopic observations of TYC\,1031\,1262\,1 were obtained with the Turkish Faint Object Spectrograph Camera (TFOSC)
attached to the 1.5 m telescope on 6\,nights (July-October, 2011) under good seeing conditions. Further details on the telescope 
and the spectrograph can be found at http://www.tug.tubitak.gov.tr. The wavelength coverage of each spectrum was 4000-9000 \AA\,in 
12\,orders, with a resolving power of $\lambda$/$\Delta \lambda$ 7\,000 at 6563 \AA~and an average signal-to-noise ratio (S/N) was 
$\sim$120. We also obtained a high S/N spectrum of the $\alpha$\,Lyr (A0 V) and HD\,50692 (G0V) for use as templates in derivation of the 
radial velocities \citep{Nidever}. 

The electronic bias was removed from each image and we used the 'crreject' option for cosmic ray removal. Thus, the resulting 
spectra were largely cleaned from the cosmic rays. The echelle spectra were extracted and wavelengths were calibrated by using Fe-Ar 
lamp source with help of the IRAF {\sc echelle} \citep{Tonry_Davis} package. 
 
The stability of the instrument was checked by cross-correlating the spectra of the standard star against each other using 
the {\sc fxcor} task in IRAF. The standard deviation of the differences between the velocities measured using {\sc fxcor} and the 
velocities in \citet{Nidever} was about 1.1 \kms.

\section{Pulsation and orbital periods}
\subsection{Pulsation period}
The pulsation period was estimated by  \citet{Ant07} and \citet{Sch09}  as 4.1523 and 4.1508\,d, respectively. Since the light variation of the pulsating 
component dominates in the light curve we first attempted to refine the pulsation period using the program PERIOD04 \citep{Len05}. A Fourier power spectrum 
of all the available V-passband data gave a spectral peak at a frequency about $f_0$=0.2409 c/d which corresponds to 4.1581\,d. This package computes 
amplitudes and phases of the dominant frequency as well as simultaneous multi-frequency sine wave fitting. Since the observations were obtained with 
different instruments in different years under dissimilar observing conditions, many systematic observational errors and computational errors affected the 
data. Therefore, we did not attempt to search for the second or the third frequencies for pulsation if they really do exist. As it will be explained in \S 4 we represented all available V-data, subtracting the eclipses, with a truncated Fourier series which includes cosine and sine  terms up to second order.  We then  calculated the light variation originated from the oscillations of the more luminous component. We separated all the data with an interval of about 6\,days and determined 
maximum times by shifting the calculated light curve along the time axis. It should be noted here that the shape of the light curve is assumed to be more or less constant during the time base of the observations.  After the best fit is obtained, the times for the mid-maximum light are read 
off directly from the observations and presented in Table\,2. Although the observations obtained by ASAS and AAVSO have relatively large scatters we had to use all the data because of limited observations both in time elapsed and the continuous observations due to its relatively longer pulsation period.    Using the ephemeris given by \citet{Ant07}
\begin{equation}
Max (HJD)=2\,453\,196.529+4^d.1523\times E 
\end{equation}
we obtained the residuals between the observed and calculated times of mid-maximum light as well as the number of the elapsed cycles. In Fig.\,2 we plot the residuals O-C(I) 
versus the epoch numbers. The variation of the residuals resembles a parabolic change, in other words TYC\,1031\,1262\,1 appears to be undergoing gradual period 
increase. A least squares solution gives the following ephemeris,

\begin{equation}
Max (HJD)=2\,453\,196.201(0.061)+4^d.15270(0.00061)\times E +9.73(2.13) 10^{-6} \times E^2
\end{equation}
The standard deviations in the last digits are given in the parentheses. In the bottom panel of Fig.\,2 the deviations from the parabolic fit are also plotted. TYC\,1031\,1262\,1 is 
undergoing period increase amounting to 2.46 ($\pm$0.54) min yr$^{-1}$.  Despite the fact that the data base is very short, over five years, the characteristics of the variation in the pulsation period could be revealed.   

\begin{figure}
\center
\includegraphics[width=10cm,angle=0]{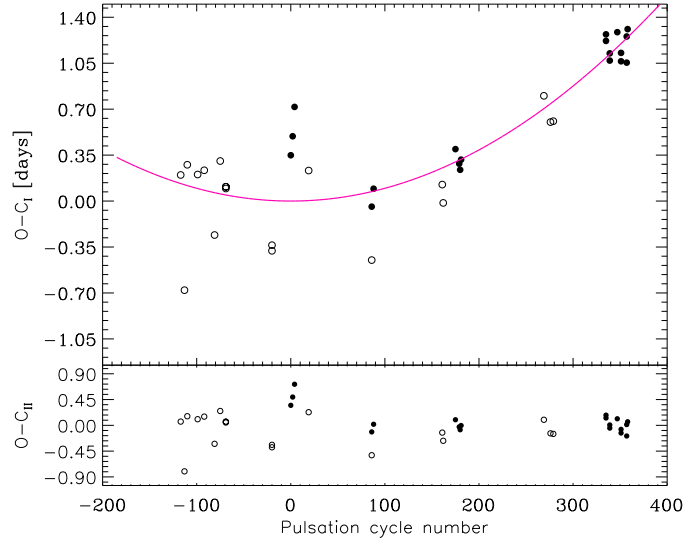}\\
\caption{The O-C residuals obtained by Eq.1 for the pulsating star of TYC\,1031\,1262\,1 are plotted against the pulsation cycle number. A parabolic fit to the data is shown by solid line. The deviations from the parabolic fit are plotted in the bottom panel. Open circles refer to ASAS and AAVSO, and the dots to the data \citet{Ant07} and this study.} 
\end{figure}

\begin{table}
\footnotesize
\begin{minipage} {12 cm}
\caption{The times of mid-maximum light for TYC\,1031\,1262\,1. The O-C(I) and O-C(II) residuals were computed with Eqs.1 and 2, respectively.}
\begin{tabular}{lccccc}
\hline
HJD-2 450 000	& E	&O-C(I)    &O-C(II) & Filter	& $ Ref.$	\\
\hline	
2710.533	&	-117&	-0.177	&	0.065	&	V	&	1	\\
2726.267	&	-113&	-1.052	&	-0.803	&	V	&	1	\\
2739.680	&	-110&	-0.096	&	0.158	&	V	&	1	\\
2785.286	&	-99	&	-0.166	&	0.107	&	V	&	1	\\
2814.387	&	-92	&	-0.131	&	0.152	&	V	&	1	\\
2859.573	&	-81	&	-0.619	&	-0.323	&	V	&	1	\\
2885.053	&	-75	&	-0.053	&	0.250	&	V	&	1	\\
2909.760	&	-69	&	-0.260	&	0.049	&	V	&	1	\\
2909.773	&	-69	&	-0.247	&	0.062	&	V	&	1	\\
3112.768	&	-20	&	-0.672	&	-0.383	&	V	&	1	\\
3112.811	&	-20	&	-0.715	&	-0.340	&	V	&	1	\\
3196.529	&	0	&	0.000	&	0.328	&	V	&	1	\\
3205.000	&	2	&	0.166	&	0.494	&	V	&	2	\\
3213.529	&	4	&	0.391	&	0.717	&	V	&	2	\\
3275.335	&	19	&	-0.088	&	0.229	&	V	&	2	\\
3552.884	&	86	&	-0.743	&	-0.521	&	V	&	1	\\
3553.291	&	86	&	-0.336	&	-0.114	&	V	&	1	\\
3561.733	&	88	&	-0.198	&	0.019	&	V	&	2	\\
3864.912	&	161	&	-0.137	&	-0.126	&	V	&	2	\\
3868.925	&	162	&	-0.277	&	-0.269	&	V	&	1	\\
3923.320	&	175	&	0.138	&	0.099	&	V	&	1	\\
3939.820	&	179	&	0.029	&	-0.026	&	V	&	2	\\
3943.926	&	180	&	-0.017	&	-0.076	&	V	&	2	\\
3948.156	&	181	&	0.061	&	-0.002	&	V	&	2	\\
4276.264	&	260	&	0.133	&	-0.297	&	V	&	2	\\
4314.080	&	269	&	0.582	&	0.099	&	V	&	3	\\
4342.949	&	276	&	0.385	&	-0.138	&	V	&	3	\\
4355.414	&	279	&	0.393	&	-0.148	&	V	&	3	\\
4388.986	&	287	&	0.747	&	0.159	&	V	&	3	\\
4588.577	&	335	&	1.027	&	0.130	&	V	&	4	\\
4588.627	&	335	&	1.077	&	0.180	&	R	&	4	\\
4605.038	&	339	&	0.879	&	-0.046	&	R	&	4	\\
4605.093	&	339	&	0.934	&	0.009	&	V	&	4	\\
4638.476	&	347	&	1.099	&	0.117	&	V	&	4	\\
4654.865	&	351	&	0.879	&	-0.132	&	R	&	4	\\
4654.929	&	351	&	0.943	&	-0.068	&	V	&	4	\\
4679.771	&	357	&	0.871	&	-0.184	&	R	&	4	\\
4679.970	&	357	&	1.040	&	0.015	&	V	&	4	\\
4684.179	&	358	&	1.127	&	0.064	&	V	&	4	\\
4684.263	&	358	&	1.211	&	0.148	&	R	&	4	\\
\hline
\end{tabular}
\end{minipage}
\begin{list}{}{}
\item[Ref:]{\small (1) ASAS, (2) \citet{Ant07}, (3) AAVSO (4) This study}
\end{list}
\end{table}

\subsection{Orbital period}
The orbital period of TYC\,1031\,1262\,1 was estimated by \citet{Ant07} as 51.38\,d. We subtracted intrinsic variations of the more luminous star from all the available data, thus, the 
remaining light variations were assumed to be originated from the eclipses and proximity effects. We revealed light variations due to the eclipses and proximity analysis with an analysis 
of these data. The shape of the primary eclipse was revealed and compared by the observations which fall in the ascending and descending branches of the eclipse. Comparing the 
computed light curve with the observations, we obtained 28 times for mid-eclipse and presented in Table \,3. The O-C(I) residuals were computed using the first ephemeris 
given by \citet{Ant07}. In Fig.\,3 we plot O-C(I) residuals versus the epoch numbers.  A linear least squares fit gives the following ephemeris,

\begin{equation}
Min (HJD)=2\,454\,699.964(0.279)+51^d.2857(0.0174) \times E
\end{equation}
The new orbital period is about 0.1 d shorter than the one estimated previously.

\begin{table}
\footnotesize
\begin{minipage} {12 cm}
\caption{The times of mid-minimum light for TYC\,1031\,1262\,1. The O-C(I) and O-C(II) residuals were computed with the ephemeris given by \citet{Ant07} and  Eq.\,3, respectively.}
\begin{tabular}{lccccc}
\hline
HJD-2 450 000	& E	&O-C(I)    &O-C(II) & Filter	& $ Ref.$	\\
\hline	
3211.7872	&	-29	&	-8.520	&	-0.892	&	V	&	1	\\
3212.8663	&	-29	&	-7.441	&	0.187	&	V	&	1	\\
3213.3491	&	-29	&	-6.958	&	0.670	&	V	&	1	\\
3211.6295	&	-29	&	-8.678	&	-1.049	&	V	&	1	\\
3213.1512	&	-29	&	-7.156	&	0.472	&	V	&	1	\\
3570.8060	&	-22	&	-6.562	&	-0.873	&	V	&	1	\\
3572.1885	&	-22	&	-5.180	&	0.510	&	V	&	1	\\
3571.5086	&	-22	&	-5.860	&	-0.170	&	V	&	1	\\
3928.4599	&	-15	&	-5.969	&	-2.218	&	V	&	1	\\
3929.3315	&	-15	&	-5.097	&	-1.347	&	V	&	1	\\
3931.9234	&	-15	&	-2.505	&	1.245	&	V	&	1	\\
3931.2237	&	-15	&	-3.205	&	0.545	&	V	&	1	\\
3931.2279	&	-15	&	-3.201	&	0.550	&	V	&	1	\\
3930.9966	&	-15	&	-3.432	&	0.318	&	V	&	1	\\
3933.5099	&	-15	&	-0.919	&	2.832	&	V	&	1	\\
4596.5385	&	-2	&	-1.003	&	-0.854	&	V	&	2	\\
4598.4954	&	-2	&	0.954	&	1.103	&	V	&	2	\\
4647.7061	&	-1	&	-0.844	&	-0.972	&	V	&	2	\\
4647.7223	&	-1	&	-0.828	&	-0.956	&	V	&	2	\\
4648.7162	&	-1	&	0.166	&	0.038	&	V	&	2	\\
4649.0758	&	-1	&	0.525	&	0.398	&	V	&	2	\\
4648.4924	&	-1	&	-0.058	&	-0.186	&	V	&	2	\\
4650.1126	&	-1	&	1.562	&	1.435	&	V	&	2	\\
4648.7959	&	-1	&	0.246	&	0.118	&	V	&	2	\\
4699.6132	&	0	&	0.054	&	-0.350	&	V	&	2	\\
4699.6096	&	0	&	0.051	&	-0.354	&	V	&	2	\\
4750.8100	&	1	&	0.242	&	-0.439	&	V	&	2	\\
4751.4869	&	1	&	0.919	&	0.238	&	V	&	2	\\
\hline
\end{tabular}
\end{minipage}
\begin{list}{}{}
\item[Ref:]{\small (1) \citet{Ant07}, (2) This study}
\end{list}
\end{table}

\begin{figure}
\center
\includegraphics[width=10cm,angle=0]{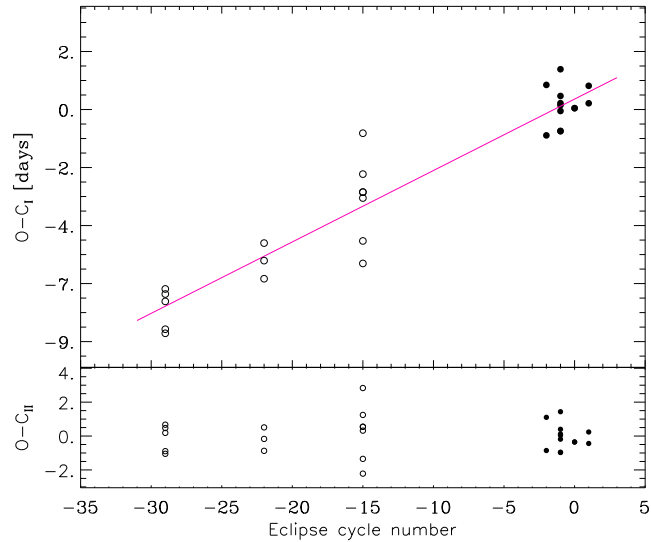}\\
\caption{The O-C residuals obtained by Eq.3 for eclipsing binary TYC\,1031\,1262\,1 and a linear least squares fit to the data. The deviations from the fit are plotted in the bottom panel. Open circles refer to the data \citet{Ant07} and the dots to this study. } 
\end{figure}

\section{Analysis}
\subsection{Effective temperature of the primary star}
We have used our spectra to reveal the spectral type of the primary component of TYC\,1031\,1262\,1. For this purpose we have degraded the 
spectral resolution from 7\,000 to 3\,000 by convolving them with a Gaussian kernel of the appropriate width, and we have also measured
the equivalent widths ($EW$) of the  photospheric absorption lines for the spectral classification. We have followed the procedures of 
\citet{hernandez}, choosing helium lines in the blue-wavelength region, where the contribution of the secondary component to the 
observed spectrum is almost negligible. From several spectra we measured $EW_{\rm He I+ Fe I\lambda 4922 }=0.81\pm 0.07$\,\AA, 
$EW_{\rm He I+ Fe I\lambda 4144 }=0.33\pm 0.05$\,\AA, $EW_{\rm Ca I\lambda 5589 }=0.35\pm 0.07$\,\AA, $EW_{\rm Ca I\lambda 6162 }=0.35\pm 0.07$\,\AA and
$EW_{\rm CH (G-band) \lambda 4300}=1.05\pm 0.02$\,\AA. From the calibration relations $EW$--Spectral-type of \citet{hernandez}, we have derived a spectral 
type of F8\,II for the more luminous star with an uncertainty of about 1 spectral subclass. In Fig.\,4 we compare spectrum of the variable, obtained on JD\,24\,55 864, with the 
spectra of some standard stars.

We have also observed the variable and comparison stars with the standard stars at the same nights. Peak to peak light and color variations of the pulsating star in V, B-V and V-R 
are about 0.45, 0.14 and 0.09 mag, respectively. The average standard magnitudes and colors of the variable are obtained as $<$V$>$ =11.64$\pm$0.04, $<$B-V$>$=0.77$\pm$0.06 
and $<$V-R$>$=0.50$\pm$0.05 mag. Comparing the location of the variable in the (B-V)-(V-R) diagram given by \citet{drill} we estimate a spectral type of F8-9\,II which is in a good 
agreement with that derived from spectroscopy. The observed infrared colors of J-H=0.368$\pm$0.033 and H-K=0.106$\pm$0.033 are obtained using the JHK magnitudes given in the 
2MASS catalog \citep{cutri}. These colors correspond to a reddened F9$\pm$2 supergiant star \citep{tok00} which is consistent with that estimated both from the spectra and BVR photometry. 

The effective temperature deduced from the calibrations of \citet{drill}, \citet{dejager},  \citet{flower},  \citet{Straiz} and \citet{vitens} are 5\,750$\pm$190\,K, 5\,740$\pm$180\,K, 5\,970$\pm$228\,K, 
6\,000$\pm$150\,K, and 5\,830$\pm$180\,K, respectively. The standard deviations are estimated from the spectral-type uncertainty. The weighted mean of the effective temperature was 
obtained for the primary star as 5\,880$\pm$165\,K. Therefore, an interstellar reddening of E(B-V)=0.21\,mag is estimated from the tables given by \citet{drill}.

\begin{figure}
\center
\includegraphics[width=10cm,angle=0]{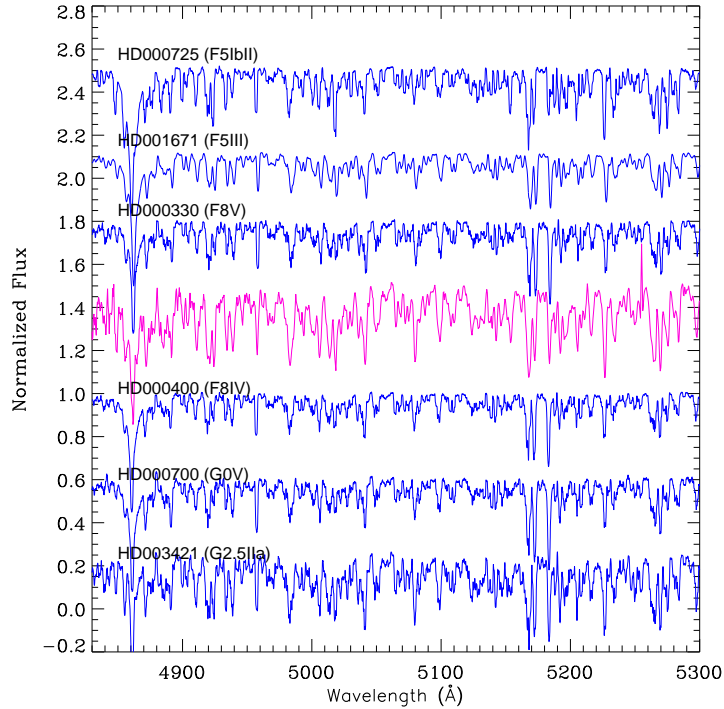}\\
\caption{Comparison of the spectrum of TYC\,1031\,1262\,1 with some standard stars with similar spectral type but different luminosity.  } 
\end{figure}

\subsection{Radial velocities}
To derive the radial velocities of the components, the 9 TFOSC spectra of the eclipsing binary  were cross-correlated against 
the spectrum of GJ\,182, a single-lined M0V star, on an order-by-order basis using the {\sc fxcor} package in IRAF. The majority of the spectra 
showed two distinct cross-correlation peaks in the quadrature, one for each component of the binary. Thus, both peaks were fitted independently
in the quadrature with a $Gaussian$ profile to measure the velocity and errors of the individual components. If the two peaks appear 
blended, a double Gaussian was applied to the combined profile using the {\it de-blend} function in the task. For each of the 9 observations we 
then determined a weighted-average radial velocity for each star from all orders without significant contamination by telluric absorption
features. Here we used as weights the inverse of the variance of the radial velocity measurements in each order, as reported by {\sc fxcor}.
We adopted a $two-Gaussian$ fit algorithm to resolve cross-correlation peaks near the first and second quadratures when spectral lines are 
visible separately. 

\begin{table}
\caption{Heliocentric radial velocities of TYC\,1031\,1262\,1. The columns give the heliocentric Julian date, the
orbital phase (according to the ephemeris in Eq.~3), the radial velocities of the two components with the 
corresponding standard deviations.}
\begin{tabular}{@{}ccccccccc@{}c}
\hline
HJD 2400000+ & Phase & \multicolumn{2}{c}{Star 1 }& \multicolumn{2}{c}{Star 2 } 	\\
             &       & $V_p$                      & $\sigma$                    & $V_s$   	& $\sigma$	\\
\hline
55747.4884	&0.4253  &4.1	&3.5   &29.7   &4.2\\
55751.3090	&0.4998  &2.0	&3.3   &---    &---\\
55796.2974	&0.3770	 &-5.9	&5.4   &45.9   &5.6 \\
55800.3794	&0.4566	 & 5.4	&3.5   &22.7   &4.8 \\   
55835.3002	&0.1375	 &-7.7	&3.6   &48.8   &3.6\\ 
55864.2074	&0.7011	 &35.8	&3.4   &-35.9  &4.8 \\
\hline \\
\end{tabular}
\end{table}

The heliocentric radial velocities for the primary (V$_p$) and the secondary (V$_s$) components are listed in Table\,4 , along 
with the dates of observations and the corresponding orbital phases computed with the new ephemeris given in previous section. The 
radial velocities are plotted against the orbital phase in Fig.\,5. The velocities in this table have been corrected for the heliocentric reference 
system by adopting a radial velocity of 14 \kms for the template star GJ\,182. The radial velocities listed in Table\,4 are the weighted averages 
of the values obtained from the cross-correlation of orders \#4, \#5, \#6 and \#7 of the target spectra with the corresponding order of the 
standard star spectrum. The weight $W_i = 1/\sigma_i^2$ has been given to each measurement. The standard errors of the weighted means have 
been calculated on the basis of the errors ($\sigma_i$) in the velocity values for each order according to the usual formula (e.g.\citet{toping}). The 
$\sigma_i$ values are computed by {\sc fxcor} according to the fitted peak height, as described by \citet{Tonry_Davis}. The observed radial velocities correspond to the pulsation phases of 0.718, 0.627, 0.360, 0.303, 0.608 and 0.484. In these phases the variation in the radius of the primary star due to oscillation is small, therefore, radial velocity changes of the Cepheid caused by the pulsation are ignored. We did not attempt to decomposition the radial velocity measurements of the primary star into the pulsation radial velocity and the orbital radial velocity.   

First we analyzed the radial velocities for the initial orbital parameters. We used the orbital period held fixed 
and computed the eccentricity of the orbit, systemic velocity and semi-amplitudes of the radial velocities. The results of the analysis 
are as follows: $e$=0.001$\pm$0.001, i.e. formally consistent with a circular orbit, $\gamma$= 11.07$\pm$0.85\kms, 
$K_1$=27.4$\pm$1.7 and $K_2$=48.1$\pm$1.7 \kms. Using these values we estimate the projected orbital semi-major
axis and mass ratio as: $a$sin$i$=76.50$\pm$2.44 \Rsun~ and $q=\frac{M_2}{M_1}$=0.570$\pm$0.041.

\begin{figure}
\center
\includegraphics[width=8cm,angle=0]{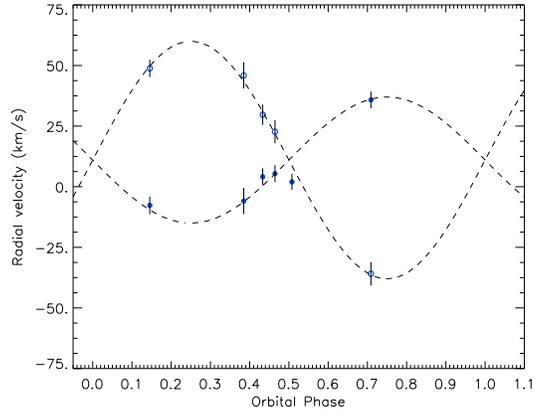}
\caption{Radial velocities folded on a period of 51.2857\,days and the model. Dots (primary) and open circles (secondary) with 
error bars show the radial velocity measurements for the components of the system } 
\end{figure}

\subsection{Intrinsic variations of the primary star}
The eclipsing  binaries provide critical information about the orbital parameters such as orbital inclination, fractional radii, luminosities, ratio of effective temperatures etc. If 
the eclipsing binary is a double-lined binary, the masses and radii of the component stars can be determined in solar units.  Using the inverse-square law one can accurately 
determine distance to the system which is independent of all distance methods.     

The observed light variations are composed of intrinsic light variations of the more luminous star and mutual eclipses. First we subtracted all the observations within the eclipses. The 
remaining observations are phased with respect to the pulsation period of 4.15270\,days. The light curves of the Cepheid show slow decline and rapid rise, i.e., asymmetric light curve 
with a sharp maximum.   The median magnitudes are taken 
as 12.408, 11.639 and 11.144\,mag for B, V and R passbands, respectively. The observed magnitudes are transformed to the flux using the median magnitudes. Then, we 
represented the intrinsic light variations of the primary star with a truncated Fourier series. Trial-and-error method showed that the observed light curves can well be 
represented by the second-order Fourier series. The coefficients are given in Table\,5 and the fits are compared with the observations in Fig.\,6.  

\begin{table}
\scriptsize
\caption{Fourier coefficients of the oscillation light curves for TYC\,1031\,1262\,1. }
\setlength{\tabcolsep}{0.8pt} 
\begin{tabular}{lrrr}
\hline
Parameters  & { $B $} & $V $ & $R $                    \\
\hline	
A$_{0}$ 								&1.0465$\pm$0.0022 					&1.0316$\pm$0.0015			&1.0016$\pm$0.0012			    \\
A$_{1}$ 								&0.2594$\pm$0.0028					&0.1948$\pm$0.0020			&0.1504$\pm$0.0016			 \\
A$_{2}$ 								&0.0543$\pm$0.0028					&0.0366$\pm$0.0020			&0.0271$\pm$0.0016			    \\
B$_{1}$ 								&-0.0019$\pm$0.0032					&0.0123$\pm$0.0023			&0.0195$\pm$0.0018			 \\
B$_{2}$ 								&-0.0530$\pm$0.0031					&-0.0371$\pm$0.0022			&-0.0288$\pm$0.0018		     \\
\hline
\end{tabular}
\end{table}

\begin{figure}
\includegraphics[width=8cm,angle=0]{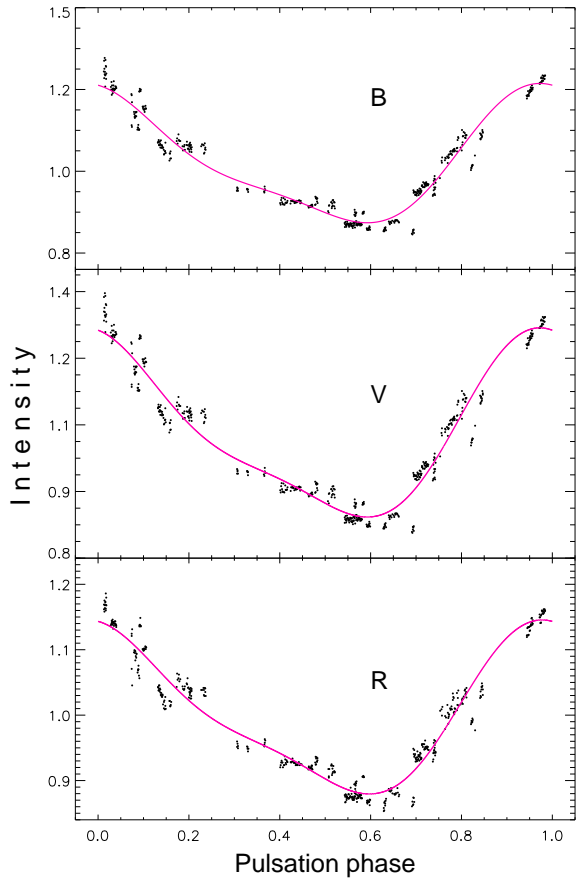}\\
\caption{ The B, V, and R passbands light curves of the pulsating primary star and their Fourier representations (solid lines) with the coefficients given in Table\,5. Note that the ordinates are normalized intensities. } 
\end{figure}

\subsection{Analyses of the light curves }
The intrinsic light variations of the primary star were computed for each oscillation phase using the coefficients given in Table\,5. After subtraction of the Cepheid light changes from the observations we obtained light variations, consisting of only from the eclipses and proximity effects. In Fig.\,7 the eclipsing light curves in the B, V and R passbands are plotted versus the orbital phases calculated by the ephemeris given in Eq.\,3. The light curve of the system with curved maxima resembles to those  $\beta$ Lyrae-type binaries. The scatter in the binary light curves is mainly caused by the variations in the pulsation period.     

We have used the most recent version of the eclipsing binary light curve modeling algorithm of \citet{wil71} (with updates, hereafter W-D), as implemented in the {\sc phoebe} 
code of Pr{\v s}a \& Zwitter (2005). It uses the computed gravitational potential of each component to calculate the surface gravities and effective temperatures. The radiative 
characteristics of the stellar disks are determined using the theoretical Kurucz atmosphere models. The code needs some input parameters, which depend upon the physical 
properties of the component stars. The U-passband observations are limited and do not cover eclipses, therefore, we excluded these observations from the analysis of light curves for orbital solution. 

The BVR photometric observations for the system  were analyzed individually. We fixed some parameters whose values were estimated from spectra, such as effective temperature 
of the hotter component and mass-ratio of the system, which are the key parameters in the W-D code. The effective temperature of the primary star has already been derived from 
various spectral type-effective temperature calibrations as 5\,880 K and the mass-ratio from the semi-amplitudes of the radial velocity curves as 0.570. A preliminary estimate for 
the effective temperature of the cooler component is made using the depths of the eclipses in the BVR passbands. Therefore, initial linear and bolometric limb-darkening coefficients 
for the primary and secondary components were taken from \citet{vanham03}, taking into account the effective temperatures and the wavelengths of the observations. The bolometric 
albedos were adopted from \citet{lucy} as 0.5, typical for a fully convective stellar envelopes. The gravity-darkening exponents are assumed to be 0.32 for  both components, because the stars are cool 
and are assumed to have convective envelopes. The rotational velocities of the components are assumed to be synchronous with the orbital one. We used Mode 2 of the W-D code which 
is for detached binaries with  no constraints on the potentials. In this mode the fractional luminosity of the secondary component is computed from the other parameters via black body 
or stellar atmosphere assumption. We assumed synchronous rotation and zero eccentricity. 
   
\begin{figure}
\center
\includegraphics[width=9cm,angle=0]{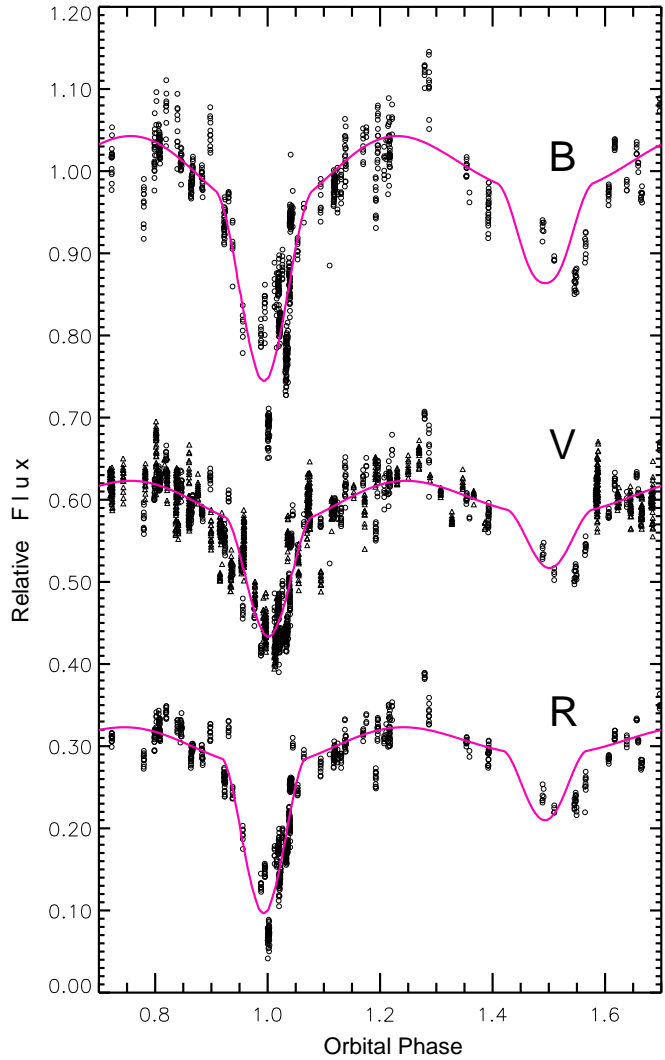}\\
\caption{The B, V and R passbands light curves, originated only from the eclipses and proximity effects,  and the computed light curves. Circles and dots are for observations obtained 
by \citet{Ant07} and in this study and solid lines are for the theoretical curves. }
 \end{figure}

The adjustable parameters in the differential correction calculation are the orbital inclination, the dimensionless surface potentials, the effective temperature of secondary, and the 
monochromatic luminosity of the hotter star. Our final results are listed in Table\,6 and the computed light curves (continuous line) are compared with the observations in Fig.\,7. The secondary minimum appears to be not very well reproduced. This is arisen mainly from the light variation of the pulsating star which is in front of the less massive secondary star at the secondary minimum. We assumed that light variation of the pulsating star is repeated with the same amplitude and the shape of its light curve is not change by time. The 
uncertainties assigned to the adjusted parameters are the internal errors provided directly by the Wilson-Devinney code. In the last three lines of Table\,6 the sums of the squares of 
residuals ($\chi^2$) , number of data points (N) and standard deviations ($\sigma$) of the observed light curves are given, respectively. Taking weight  inversely proportional to 
$\sigma$, the adopted parameters given in the last column of Table\,6 are obtained.

\begin{table}
\scriptsize
\caption{Results of the BVR light curves analyses for TYC\,1031\,1262\,1. }
\setlength{\tabcolsep}{0.8pt} 
\begin{tabular}{lrrrr}
\hline
Parameters  & { $B $} & $V $ & $R $   & Adopted                      \\
\hline	
$i^{o}$										&74.85$\pm$0.13					&71.26$\pm$0.39			&76.31$\pm$0.49		&74.0$\pm$0.4  \\
T$_{eff_1}$ (K)								&5880[Fix]						&5880[Fix]				&5880[Fix]			&5880[Fix]    \\
T$_{eff_2}$ (K)								&5073$\pm$102					&4907$\pm$72			&4661$\pm$83		&4840$\pm$80			   \\
$\Omega_1$									&3.416$\pm$0.037			    &3.594$\pm$0.026		&3.709$\pm$0.038	&3.604$\pm$0.033		       \\
$\Omega_2$									&4.118$\pm$0.182			    &4.054$\pm$0.069		&4.617$\pm$0.100	&4.292$\pm$0.104			   \\
r$_1$				   						&0.3615$\pm$0.0051			    &0.3380$\pm$0.0032		&0.3258$\pm$0.0042	&0.3378$\pm$0.0040	       \\
r$_2$										&0.1760$\pm$0.0101  			&0.2004$\pm$0.0049		&0.1839$\pm$0.0060	&0.1890$\pm$0.0064	       \\
$\frac{L_{1}}{(L_{1}+L_{2})}$ 	            &0.903$\pm$0.015				&0.882$\pm$0.011		&0.892$\pm$0.012	               \\
$\chi^2$									&2.532							&2.186					&0.653					     	   \\		
N									        &972							&2557					&972					     	   \\
$\sigma$									&0.0512							&0.0293					&0.0260					     	   \\
\hline
\end{tabular}
\end{table}

\section{Results and Discussion}
A comparison of the results obtained in three light curves  reveals some differences in the inclination angle and especially in the fractional radii of 
the stars. The B and R light curves are constructed by only our own observations. The scatter in the B-passband observations is significantly larger compared to the other bands due to its faintness.  Combining the spectroscopic results along with the photometric solutions, listed in the last column of Table\,6, the 
absolute masses, radii, luminosities and surface gravities of the stars are obtained and presented in   Table\,7. The mass and radius of the pulsating component were determined for the first time directly from radial velocities and multi-color light curves. They are unexpectedly large compared to the Type\,II Cepheids.
 The luminosity and absolute bolometric magnitude M$_{bol}$ of each star were computed from their effective temperatures and radii. The  bolometric magnitude and effective temperature for the Sun are taken as 4.74 mag and 5770 K (\citet{drill}). Comparison of its location in the Hertzsprung-Russell diagram (HR), constructed for the metal-poor cepheids by \citet{gin85}, their Fig.\,1, indicates that TYC\,1031\,1262\,1 is in the instability strip of the Type\,II Cepheids, about 20 times more luminous than the RR \,Lyrae stars . It seems to have the same luminosity and effective temperature with the Type\,II Cepheids having pulsation periods of about 10 days, closer to the blue edge of the IS. When we compare with the solar composition models, for example \citet{bar98}, the primary star appears to be an evolved 5 M$_{\odot}$ star and the secondary is consistent with an evolved 3 M$_{\odot}$ star. 

After applying bolometric corrections to the bolometric magnitudes, we obtained absolute visual magnitudes of the components. Taking into account the light contributions 
of the stars and total apparent visual magnitude we calculated their apparent visual magnitudes. The light contribution of the secondary star $L_2$/($L_1$+$L_2$)=0.02, 0.04 and 0.08 are obtained directly from the B-, V-, and R-bandpass 
light curve analysis, respectively. This result indicates that the light contribution of the less massive component is very small, indicating its effect on the 
color at out-of-eclipse is almost negligible for the shorter wavelengths. For the bolometric corrections given by (\citet{drill}) we 
calculated the distance to TYC\,1031\,1262\,1 as $d$=5074$\pm$207 pc. However the infrared JHK magnitudes give a distance of 5040$\pm$255 pc. Estimating 
distances to the stars are strongly depended on the bolometric corrections. If we adopt the bolometric corrections given by \citet{cod76} and \citet{bes98}, the distances to TYC\,1031\,1262\,1 are obtained as 5082$\pm$540, and 5110$\pm$650 pc, respectively. The average distance to the system was obtained as 
5070$\pm$250 pc. The distance from the galactic plane is calculated as 970 pc using its galactic latitude of 11 degrees, about 14 times larger than the mean scale height of the Type\,I Cepheids. We have calculated Galactic space-velocity components of the system using the coordinates, distance, systemic radial velocity, and proper motions and listed in the last line of Table\,7. The U, V, and W are velocities toward the galactic center, toward the direction of galactic rotation and toward the north galactic pole, respectively. With planer and vertical eccentricities of 0.50 and 0.26 and the velocity components the system is most probably belonging to the thick-disk population. This result confirms the earlier suggestion of \citet{har84} that many field  Type\,II Cepheids are metal rich and have kinematic  properties similar to the old-disk or thick-disk population.

The pulsation period of TYC\,1031\,1262\,1 has shown a secular period increase with an amount to 2.46 ($\pm$0.54) min yr$^{-1}$. \citet{vin93} investigated secular period change in AU\, Peg, another Type\,II Cepheid which belongs to a binary system. The observed rate of the period increase in TYC\,1031\,1262\,1 is about 5 times larger than that of AU\, Peg. \citet{mcc97} observed a decrease in the pulsation period of V19 in the globular cluster  NGC\,5466 which is known as an AC with a mass of 1.66 M$_{\odot}$ and a pulsation period of 0.82 d. They propose that this would be an indication of blueward evolution. In contrary, TYC\,1031\,1262\,1 shows a period increase and its large magnitude should indicate faster evolutionary timescale redward.  \citet{San95} and \citet{Die96} proposed that all of the BL\, Her Cepheids appear to have increasing periods and are consistent with post-blue horizontal branch models of Population\, II stars. The observed period increase in TYC\,1031\,1262\,1 appears to confirm this suggestion. Unlike the BL\, Her Cepheids, the TYC\,1031\,1262\,1 has a companion very close to it. Period change rates are determined for only a few ACs. Yet the cause of such a relatively large pulsation period change in a Type\,II Cepheid has not been obvious. Observations covering a long time span are crucially needed for better understanding of such a large increase rate in pulsation period in the TYC\,1031\,1262\,1, as well as in other  Type\,II Cepheids.

\begin{table}
 \setlength{\tabcolsep}{2.5pt} 
  \caption{Fundamental parameters of TYC\,1031\,1262\,1.}
  \label{parameters}
  \begin{tabular}{lcc}
  \hline
  & \multicolumn{2}{c}{\hspace{0.25cm} TYC\,1031\,1262\,1} 																														\\
   Parameter 												& Primary	&	Secondary					\\
   \hline
   Spectral Type											& F8($\pm$1)II  	&G6($\pm$1)II    				\\  
   Mass (M$_{\odot}$) 								& 1.640$\pm$0.151       &0.934$\pm$0.109			\\
   Radius (R$_{\odot}$) 								& 26.9$\pm$0.9    &15.0$\pm$0.7				\\
   $T_{eff}$ (K)											& 5\,880$\pm$200&4\,890$\pm$125    	\\
   Luminosity (L$_{\odot}$) 								& 764$\pm$144   & 109$\pm$26				\\
   Gravity($cgs$) 										& 62$\pm$3          &114$\pm$11	                \\
   $a$ (R$_{\odot}$)									&\multicolumn{2}{c}{79.58$\pm$2.54}			\\
   $V_{\gamma}$ (km s$^{-1}$)				&\multicolumn{2}{c}{11.07$\pm$0.85} 				     	\\
   $i$ ($^{\circ}$)										&\multicolumn{2}{c}{74.0$\pm$0.4	} 			\\
   $q$															&\multicolumn{2}{c}{0.570$\pm$0.041} 	\\   
   $d$ (pc)													& \multicolumn{2}{c}{5070$\pm$250}			\\
	$\mu_\alpha cos\delta$, $\mu_\delta$(mas yr$^{-1}$) & \multicolumn{2}{c}{2.7$\pm$0.8, 0.8$\pm$0.8} 		\\
	$U, V, W$ (km s$^{-1}$)  						& \multicolumn{2}{c}{-14.5$\pm$13.0, 47.7$\pm$14.9, -47.2$\pm$19.1}	\\ 
\hline  
 \end{tabular}
\end{table}

As it is quoted in \S 1, Type\,II Cepheids are divided into three subclasses:  BL\,Her, W\,Vir, and RV\,Tau. All three classes are characterized by the presence of Balmer emission, especially H$\alpha$, during some parts of their pulsation cycle. Evolutionary scheme of these classes is suggested by \citet{gin85}. In this scheme the BL\,Her stars are evolving from horizontal branch towards the lower asymptotic giant branch. However, the W\,Vir variables are on loops to the blue from the asymptotic giant branch. The RV\,Tau stars are moving to the blue in a post-asymptotic giant branch phase. He investigates evolutionary status of Type\,II Cepheids as post-horizontal-branch stars and calls attention to a different structure and evolution of the ACs. He estimates masses of Type\,II Cepheids around 0.6 M$_{\odot}$ with a radius of about  8 R$_{\odot}$ in the post-horizontal-branch phase. \citet{bono97} suggest masses between 0.52 and 0.59 M$_{\odot}$ for the Type\,II Cepheids using their pulsation properties rather than evolutionary tracks. Now there is a consensus that Type\,II Cepheids  are fundamental pulsators with masses below 0.8 M$_{\odot}$. 

With a mass of about 1.640 M$_{\odot}$, luminosity of 760 L$_{\odot}$ and radius of 27 R$_{\odot}$ the pulsating primary component of TYC\,1031\,1262\,1 cannot be a Type\,II Cepheid. If it were a Type\,I Cepheid its mass would be greater than 4 M$_{\odot}$ according to the models given by \citet{bono00} for Z=0.02 and even for Z=0.004. However the ACs are usually accepted to be metal-poor horizontal branch stars with masses above 1.5 M$_{\odot}$. \citet{sos08} presented 83 ACs in the Large Magellanic Cloud and showed that ACs are located between Type\,I and Type\,II Cepheids in the P-L diagram.  They are found in every dwarf spheroidal galaxy, and unlike Type\,II Cepheids, are absent in globular clusters, except a few ACs in  ${\omega}$ Cen. \citet{dem75} propose for their origin that they are young single stars due to recent star formation. In contrary, \citet{ren77} suggest that they are formed as a consequence of mass transfer in binary systems with the same age of the stellar systems they belong. \citet{bon97} examined evolution and pulsation properties of ACs and concluded that models of masses between 1.5 and 2.2 M$_{\odot}$ fit most of the stars. Concerning the origin of ACs, as single young stars or old binary systems, they have noted that their results did not support a single interpretation.  The mass, luminosity, kinematic properties and light curves of TYC\,1031\,1262\,1 fulfil most of the properties of the Anomalous Cepheids. As pointed out by \citet{sos08} and \citet{fio12} the mode identification of an AC is very complex and can not be based only on their light curves. However, \citet{sos08} compared light curves of the ACs in the LMC. They proposed that the ACs pulsating in the first overtone have generally smoother light curves than for the fundamental-mode pulsators with rounded maxima and minima. The shape of the light curve of the pulsating star looks like to those of fundamental-mode pulsators. If we compare its location on the M$_{V}$\,-\,$log$\,P plane given by \citet{fio12} for the ACs in the LMC, whose pulsating periods are shorter than 2.4 d,  the pulsating star is located on the extension of fundamental - pulsators, as though its classification as a fundamental-pulsator is supported. In Fig.\,8 we compare position of TYC\,1031\,1262\,1 on the color-magnitude diagram with other ACs taken from \citet{harr84} and \citet{nem94}. It is located among the brightest ACs close to the blue edge, with a luminosity of about 800 times solar. The binary Cepheid ST\,Pup with a pulsation period of about 18.47 d, is located very close to the TYC\,1031\,1262\,1. In contrary, the binary Cepheids TX\,Del with a pulsation period of 6.2\,d and AU\,Peg with a period of 2.4\,d, are located at the red-edge of the IS, with lower luminosity.

\begin{figure}
\center
\includegraphics[width=9cm,angle=0]{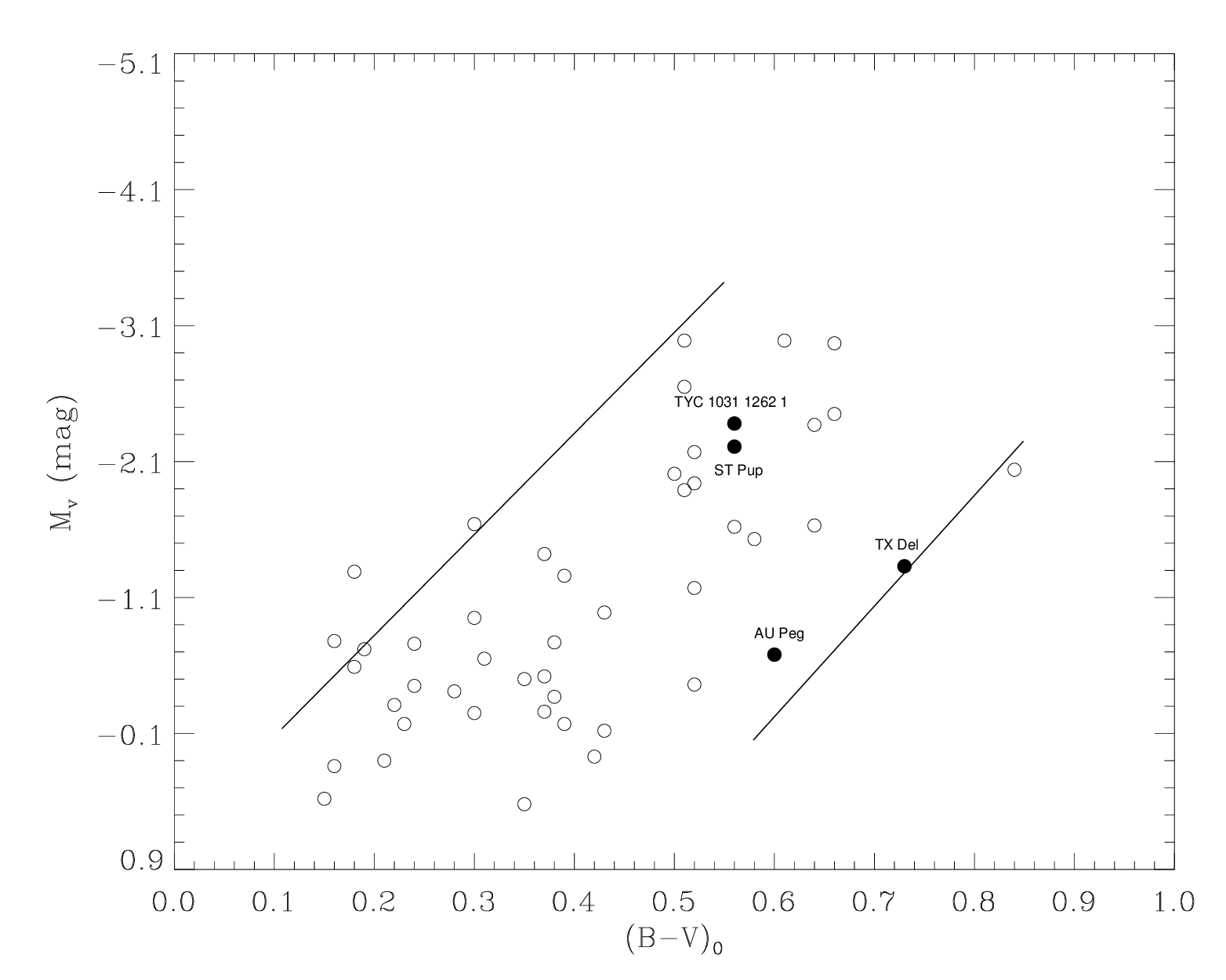}
\caption{The position of TYC\,1031\,1262\,1 on the color-magnitude diagram. The circles refer to the Type\,II Cepheids in globular clusters and the dots for TYC\,1031\,1262\,1, ST Pup, AU Peg and TX Del. The lines show the borders of instability strip for Type\,II Cepheids.}
\end{figure}

\section{Conclusions}
We have obtained multi-color light curves and radial velocities of the eclipsing binary system TYC\,1031\,1262\,1 of which the luminous component is a Cepheid. The astrophysical parameters 
of the component stars were obtained directly by analysing the light curves and radial velocities. In earlier studies the brighter star was classified as a metal-poor Type\,II Cepheids. In contrary, some 
investigators include it into the classical cepheids. The mass of the pulsating primary star with an amount of 1.64 M$_{\odot}$ indicates that its structure and evolution are different from those of 
the RR\,Lyrae and W\,Vir stars. The galactic velocity components $U, V, W$ have been determined as -14.5, 47.7 and -47.2 km s$^{-1}$ and the distance from the galactic plane as about 970 pc. All 
these properties and the asymmetric pulsating light curve lead us to classify the Cepheid as an Anomalous Cepheid. The pulsation period appears to be increasing with an amount of 2.46 ($\pm$0.54) 
min yr$^{-1}$, the largest period change observed in the Type\,II Cepheids so far. This observed increase rate indicates that the star evolves redward in the IS with shorter timescales. 

The pulsating primary star fills almost 85 per cent of its corresponding Roche lobe while the less massive star 61 per cent. The pulsating component in  TYC\,1031\,1262\,1 is very close to its Roche 
lobe, therefore, mass loss or transfer to its companion has taken place. Both components are supergiant and the separation is only two times the sum of their radii. The companion should 
unexpectedly affect both pulsation and evolution of the Cepheid. However, it is not clear how such a close companion can cause any period variation on pulsation by tidal interaction alone. Most 
of the ACs are faint stars and have relatively longer orbital periods, therefore, measurements of the radial velocities ACs in binary systems are very difficult. In addition the velocity variations due to the 
orbital motion may be affected by the pulsation depending on the inclination of the orbits. In the case of TYC\,1031\,1262\,1 the orbital inclination is sufficiently large to separate the orbital velocities of 
both components. If mass loss or transfer from the pulsating star has taken place, the orbital period of the system would be changed, which can be revealed by the observations to be made in coming 
years. This will yield us some clues about the further evolution of the ACs in binary systems.                  

\section*{Acknowledgments}
Special thanks goes to Dr. S. Bilir and F. Soydugan for his valuable suggestions and discussion about the kinematic properties of the system. 
We thank to T\"{U}B{\.I}TAK National Observatory (TUG) for a partial support in using RTT150 telescope with project numbers 
10ARTT150-483-0, 11ARTT150-123-0 and 10CT100-101. 
We also thank to the staff of the Bak{\i}rl{\i}tepe observing station for their warm hospitality. This study is partly supported by Turkish Scientific and Technology Council under project number 108T237.
%
%
The following internet-based resources were used in research for this paper: the NASA Astrophysics Data System; the SIMBAD database operated at CDS, Strasbourg, France; T\"{U}B\.{I}TAK 
ULAKB{\.I}M S\"{u}reli Yay{\i}nlar Katalo\v{g}u-TURKEY; and the ar$\chi$iv scientific paper preprint service operated by Cornell University. The authors are indebted to the anonymous referee 
for his/her valuable suggestions which improved the paper.

\bibliographystyle{mn_new}

\label{lastpage}

\end{document}